\documentclass[12pt,eqsecnum,preprint]{aastex}

\begin{document}

\title{The Velocity Field of Quasar Broad Emission Line Gas}
\author{Brian Punsly\altaffilmark{1}} \altaffiltext{1}{4014 Emerald Street No.116, Torrance CA, USA 90503 and
International Center for Relativistic Astrophysics,
I.C.R.A.,University of Rome La Sapienza, I-00185 Roma, Italy}

\begin{abstract}
In this Letter, the broad emission line (BEL) profiles of
superluminal quasars with apparent jet velocities, $\beta_{a}>10$,
(ultraluminal QSOs, or ULQSOs hereafter) are studied as a diagnostic
of the velocity field of the BEL emitting gas in quasars. The ULQSOs
are useful because they satisfy a very strict kinematical
constraint, their parsec scale jets must be propagating within
$12^{\circ}$ of the line of sight. We know the orientation of these
objects with great certainty. The large BEL FWHM, $\sim
3,000\,\mathrm{km/s} - 6,000\,\mathrm{km/s}$, in ULQSOs tend to
indicate that the BEL gas has a larger component of axial velocity
(either random or in a wind) along the jet direction than previously
thought.
\end{abstract}


\section{Introduction}The most prominent feature of quasar
optical/UV spectra are the conspicuous broad emission lines (BELs).
The large line widths indicate a large spread in the velocity of the
emitting gas. The nature of the velocity field of the BEL region is
not well-constrained because the line of sight to the gas is not
well defined in general. It is impossible to extricate the effects
of orientation on the BEL FWHM (full width half maximum) from the
effects of the central black hole mass and the luminosity of the
accretion disk. The geometrical orientation of BLR (broad line
region) and its relationship to the FWHM has received tremendous
attention due to the potential consequences for studies of virial
black hole estimates.
\par Previous efforts to find orientation
information on the BLR relied on one of the following methods:
segregating flat spectrum radio loud quasars, FSQSRs, from steep
spectrum radio loud quasars, SSQSRs \citep{cor97}; correlating the
FWHM with the radio spectral index, $\alpha$ \citep{jar06};
correlating FWHM with the core dominance parameter of radio loud
quasars, R (the rest frame flux density ratio at 5 GHz between the
unresolved core and resolved radio emission), \citet{bro86}; or with
$R_{V}$ of \citet{wbr95} (the ratio of the radio core flux density
at 5 GHz to the optical flux density which is isotropic, if it is
from an accretion disk); the rotation of the polarization vector
across the BEL can constrain the scattering geometry has been
applied to some Seyfert 1 galaxies and supports a disk-like
flattened geometry of the BLR \citep{smi05}.
\par These methods give
vague orientation information which has limited our ability to
resolve the geometry of the BEL velocity field. In particular,
SSQSRs are probably concentrated in the range of $\theta <
55^{\circ}$ lines of sight from the jet axis based on the statistics
of the complete 3CRR sample \citep{ogl06}. Due to relativistic
Doppler boosting, the flat spectrum radio core flux, if viewed at
$\theta < 20^{\circ}$, will be drastically enhanced so as to swamp
the steep lobe flux at 5 GHz and the source will appear as a FSQSR
\citep{lin85}. Thus, for SSQSRS, we expect $20^{\circ}<\theta <
55^{\circ}$. The FSQSRs, are restricted to lines of sight within
$\theta= 30^{\circ}$ from the jet axis. This constraint on the
FSQSRs is much weaker than is generally acknowledged and provides
tremendous overlap between the two sub-populations. We illustrate
this with an example. Recall that for an unresolved source the
observed flux density, $S_{\nu}(obs)$, is Doppler enhanced relative
to the intrinsic luminosity,
$S_{\nu}(obs)=\delta^{3+\alpha}S_{\nu}(intr)$, where the Doppler
factor, $\delta$ is given in terms of $\Gamma$, the Lorentz factor
of the outflow; $c\beta\equiv \sqrt{v_{x}^{2} + v_{y}^{2} +
v_{z}^{2}}$, the three velocity of the outflow; and $\alpha$, is the
spectral index: $\delta=1/[\Gamma(1-\beta\cos{\theta})]$
\citep{lin85}. Consider an FR II quasar that is viewed at
$\theta=50^{\circ}$. It is a "typical" powerful FR II quasar at
z=0.5, with steep spectrum lobes $S_{\nu}(obs)$=3 Jy at 178 MHz and
the lobe has $\alpha=1$. The radio core is chosen to have 25 mJy at
5 GHz with $\alpha=0$. Then $\alpha=0.85$ between 2.7 GHz and 5 GHz,
a very typical value for a SSQSR. If the jet is moving with
$\beta=0.995$, ($\Gamma=10$) and the same object were viewed at
$\theta=30^{\circ}$ then the observed core flux density is about 500
mJy at 5 GHz and the lobe flux is only about 100 mJy at 5 GHz. The
same object that was a typical lobe dominated SSQSR viewed at
 $\theta=50^{\circ}$ is transformed into a typical FSQSR ($\alpha=0.23$ between 2.7 GHz
and 5 GHz) if it is viewed at $\theta=30^{\circ}$. This example
indicates that FSQSRs should exist for lines of sight
$\theta=30^{\circ}$ from the jet axis which overlaps the range of
$\theta$ for SSQSRs (based on the angular range discussed above, the
average value for the lines of sight to SSQSRs is
$\theta=39^{\circ}$)The difference between $30^{\circ}$ and
$12^{\circ}$ (which is the maximum $\theta$ of the ULQSO sample
defined below) is highly significant for this analysis because the
projection of an equatorial velocity component onto the line of
sight $\sim \sin{\theta}$ (c.f. equations (3.1) and (3.2)): this
equates to a factor of 2.5 for the two values of $\theta$. It is
concluded that segregating FR II QSOs by spectral index gives very
crude orientation information.
\par The quantities R and $R_{V}$
give improved angular information because it captures the Doppler
enhancement factor. However, this is offset by the fact that
\textbf{intrinsic} values of radio core luminosity, radio lobe
luminosity and accretion disk luminosity are not "standard candles,"
but vary considerably from quasar to quasar. Secondly, there is
large optical contamination of the accretion disk luminosity from
the Doppler boosted jet in many FSQSRs that further smears the
angular resolution of this method. The correlation analysis of
\citet{bro86,wbr95} is convincing, but it really only tells us that
there is some component of the BLR that is concentrated near the
equatorial plane, it doesn't tell us if this represents the bulk of
the BEL emitting gas or a much smaller fraction like 1/3 of the
total BEL gas (cf. eqn (2.2)), the correlation could exist in either
case.
 \par Our understanding of the velocity field of the BEL gas has been limited
by the poor orientation resolution of these methods. In the end, one
is left with a large uncertainty in the orientation factors that are
relevant for virial black hole estimates \citet{kro01}. Thusly
motivated, in this paper, we study a sample of objects in which the
line of sight is very well-constrained. Even though the sample is
not large, 76 broad lines from 33 QSOs, the accurate orientation
information is very insightful. The sample consists of ultraluminal
quasars (ULQSOs, hereafter) that are defined by the condition that
their projected apparent velocity on the plane of the sky is,
$\beta_{a}\geq 10$. Since,
$\beta_{a}=(\beta\sin{\theta})/(1-\beta\cos{\theta})$, the
$\beta_{a}\geq 10$ condition and the fact the intrinsic velocity of
the jet plasma is subluminal $\beta<1$, means that the line of sight
makes an angle $\theta< 12^{\circ}$ to the jet axis. With the
knowledge that we are viewing the quasars extremely close to the
polar axis, implies that the FWHM of the BELs will give unambiguous
information on Doppler shift associated with one component of the
velocity field of the BEL emitting gas, the axial component along
the jet axis. This article considers all of the prominent BELs that
were accessible to observation, H$\beta$, MgII, CIII] and CIV. There
is strong blending between  CIII] and SiIII, thus the analysis of
these lines is not treated with as much weight as the other BELs in
the following. The FWHM that are found, correspond to velocities
that average $\approx 4,000$ km/sec for all BELs. The implication is
that QSOs have prodigious quantities of high velocity gas moving
both parallel and \textbf{antiparallel} to lines of sight along the
jet direction. Formally, the results of this paper are restricted to
radio loud AGN. The extent to which it applies to radio quiet AGN is
unclear.
\section{The Ultraluminal Quasar and Comparison Samples}Fortunately, there are
three large multi-epoch VLBI studies. The sample in table 1
represents all the $\beta_{a}\geq 10$ AGN that were found by
\citet{kel04,jor01,ver94} for which there is some archival FWHM
data. There are many preferences for de-constructing BELs into
various components \citep{mar03,ves06}. There is likely much merit
in this endeavor, but the decompositions and FWHM estimates are not
produced uniformly in the references cited in table1 and table 2
(described below). The choice is to either greatly limit the sample
size to a few sources from one reference or introduce scatter
depending on whether narrow components were subtracted or how the
very broad component was fit. A very small sample of ULQSOs is not
of much interest from a statistical point of view. The FWHM in the
control samples in table 2 are drawn from largely the same
references that appear in table 1. Thus, the degree of scatter from
different derivation techniques occurs in all samples and is likely
to be quantitatively similar. In order to minimize this scatter, an
effort was made to find single Gaussian fits to the data, if they
were available. This gives us the maximum amount of commonality with
the control samples in table 2 for the low ionization BELs because
many of these FWHM come from single Gaussian fits: \citet{law96},
for the ULQSOs; \citet{law96,cor97} for the sample of \citet{guu01};
and \citet{aar05,law96} for the 3CRR sample. It would be important
to followup this work with a uniform reanalysis of the raw data that
segregates the various components for comparison. This is far beyond
the scope of this effort.
\par In order to obtain more information than just the
fact that there are large BEL velocities along the jet axis, it is
of interest to compare this data with samples that represent sources
viewed from far more oblique angles. Unfortunately, we can not use
the large database of radio quiet quasars because there are numerous
claims in the literature that the BELs of radio quiet and radio loud
quasars have significant differences. For example, some authors
claim that H$\beta$ FWHM is larger in radio loud QSOs \citep{mcc01}.
Thus, we must restrict our discussion to radio loud objects and the
SSRQRs are the sources that are likely viewed at relatively steep
angles to the line of sight. Given that we are dealing with a
relatively small number of sources $\sim 20$ for a particular ULQSO
BEL and samples of SSQSRs in the literature are of modest size $\sim
50$, it is unrealistic to match all parameters (i.e., redshift, UV
luminosity, extended radio power). First of all, it is impossible to
match UV power appropriately because the accretion disk emission is
contaminated by the jet optical emission in the ULQSOs. It is also
extremely difficult to match extended radio power, since ULQSOs are
core dominated and one needs very deep, high dynamic range radio
maps (to find the relatively faint extended flux) which do not exist
except for a few of the ULQSOs. Thus, we chose to match the samples
by redshift since this is often related to one or more of these
other parameters by selection biases. The samples used for
comparison and contrast are listed in table 2 and the distributions
of FWHM can be directly compared. As a check of our sample selection
method, we can compare to the carefully selected samples of
\citet{jar06}. They find for a combination of H$\beta$ and MgII,
FWHM(FSQSR)/FWHM(SSQSR)=0.77, and we find for H$\beta$;
FWHM(ULQSO)/FWHM(SSQSR)=0.67, for the combined SSQSR sample of
\citet{guu01,cor97}, which is consistent with the ULQSOs being
viewed a little closer to the pole than a typical FSQSR as discussed
in the introduction and verifies the lack of a significant selection
bias to our study.
\par First we compare H$\beta$ to the sample of SSQSRS from
\citet{guu01} as was done in \citet{osh02} in their comparison with
FSQSRS. The sample is comprised of all the SSQSRS with $>1$ Jy at 5
GHz from the S4 and S5 surveys and is 97\% complete \citep{guu01}.
This is well matched in redshift to the ULQSOs, but it is small. So,
we construct a larger H$\beta$ control sample by adding the
\citet{cor97} sample of SSQSRS which is at similar redshift. The
essential conclusions of this article are unaffected by which sample
is used as the control. There was no existing list of MgII FWHM for
SSQSRs, so for a comparison the noncompact ($> 20$ kpc ) SSQSRs from
the 3CRR catalog that are known not to be dominated by a powerful
blazar core were chosen to represent the steep lines of sight. The
samples for CIII] and CIV are drawn from the same references, the
high z sources are from \citet{cor94} and the low redshift
contribution is from the HST data of \citet{wil95} in the tables.
\par From table 2, based on the size of
the standard deviation, from a statistical point of view one can not
claim that there are differences between the SSQSR FWHMs and the
ULQSO FWHMs for any of the four BELs. However, we note that the
means and medians for H$\beta$ and MgII FWHMs are smaller for the
ULQSOs and the opposite is true for CIII] and CIV. We consider the
popular notion that the BEL gas is distributed in an equatorial
pancake, orthogonal to the jet axis, with a random velocity,
$v_{r}$, superimposed on an equatorial velocity, $v_{p}$, that is
predominantly bulk motion from Keplerian rotation \citet{jar06},
\begin{eqnarray}
&& FWHM \simeq 2\sqrt{v_{r}^{2}+v_{p}^{2}\sin^{2}{\theta}}\;.
\end{eqnarray}
In \citet{lab06}, they assume that the random component is from
Keplerian orbits of random inclination and find the relation
\begin{eqnarray}
&& FWHM =
2(c_{1}\frac{1}{\sqrt{2}}\sin{\theta}+c_{2}\frac{1}{\sqrt{3}})
v_{BLR}\;,
\end{eqnarray}
where $c_{1}$ and $c_{2}$ "represent the relative fraction of the
FWHM due to the" planar and isotropic components, respectively.
Using the angular coordinate ranges motivated in the Introduction
($0^{\circ} < \theta < 12^{\circ}$ for ULQSOs and $20^{\circ} <
\theta < 55^{\circ}$ for SSQSRs) and the mean FWHM from Table 2, one
can estimate the ratio $v_{r}/v_{p}$ for each BEL after
appropriately averaging (3.1). For example, for H$\beta$, we solve
the equation,
\begin{eqnarray}
&&
\frac{1}{\cos{0^{\circ}}-\cos{12^{\circ}}}\int^{12^{\circ}}_{0^{\circ}}\sqrt{v_{r}^{2}+v_{p}^{2}\sin^{2}{\theta}}\,
\sin{\theta}\, d\theta \nonumber\\
&&
=\left(\frac{3898}{5836}\right)\left(\frac{1}{\cos{20^{\circ}}-\cos{55^{\circ}}}\right)\int^{55^{\circ}}_{20^{\circ}}\sqrt{v_{r}^{2}+v_{p}^{2}\sin^{2}{\theta}}\,
\sin{\theta}\, d\theta\;.
\end{eqnarray}
This equation can be solved numerically to give, $ v_{r}/v_{p}=0.537
$ for H$\beta$. Similarly, for Mg II, $ v_{r}/v_{p}=0.502 $ and for
CIV (and CIII), the only consistent solution is $v_{p}=0$. If the
mean values of the FWHM from table 2 are replaced in (3.3) by the
median values, the results are slightly changed; for H$\beta$,
$v_{r}/v_{p}=0.563 $; for Mg II, $ v_{r}/v_{p}=0.496 $; and for CIV
(and CIII), the only consistent solution is $v_{p}=0$. This makes
the result seem more robust. If one assumes that $ v_{r} $ and $
v_{p} $ represent the rms velocity of the two BEL components then
one can claim approximate equality in (2.1), as in
\citet{mcc01,lam06}. Using the mean values in table 2 transforms the
average of (2.1) over solid angle in (2.3) into two equations in two
unknowns and one can solve for the mean values of $ v_{r} $ and $
v_{p} $: $ v_{r}= 1877 $ km/ s, $ v_{p}= 3496 $ km/s for H$\beta$;
$v_{r}= 1979 $ km/ s, $ v_{p}= 3942 $ km/s for MgII; and $v_{r}=
2150 $ km/ s, $ v_{p}\approx 0 $ km/s for CIV. Similarly, one can
average (2.2) over solid angle, to find $c_{1}/c_{2}=1.04$ for
H$\beta$ and $c_{1}/c_{2}=0.93$ for MgII and $c_{1}/c_{2}=0$ for
CIV.
\par The improved angular resolution of this study can be compared
to previous analysis of the BLR geometry. From (2.1) and (2.2), the
low ionization BLR can be characterized by a mixture of planar and
random components. For CIV, the model breaks down and the only
consistent solution is purely random motion within the model. The
value of $v_{r}$ is larger than has what has been estimated by lower
angular resolution studies. In section II b of \citet{bro86}, it is
estimated that $0.15 < v_{r}/v_{p}< 0.31$ and in \citet{cor97}
$v_{r}/v_{p}\approx 0.15$ for H$\beta$, compared to our value of
0.537. Similarly, \citet{nik06} assume that $(v_{r}/v_{p})^{2}\ll
0.28$, while we find to the contrary that $(v_{r}/v_{p})^{2}\approx
0.3$. Furthermore, the ULQSO analysis applied to CIV strongly
contradicts the findings of \citet{lab06} who assert that the CIV
FWHM in quasars are consistent with a planar distribution of BEL gas
near the equator. The discrepancy could be a manifestation of the
difference between their radio quiet sample and our powerful jet
sample. This effort seems to be the first that actually performs the
numerical averaging of (2.1).
\section{Conclusion} In this paper, we study the FWHM of the BELs
of ULQSOs. We applied our findings to the planar models of the BLR,
focusing on the velocity components of the BEL gas both redward and
blueward along the jet axis (perpendicular to the inner accretion
disk). Our analysis of the ULQSOs supports the notion that the high
ionization BELs come from gas with a different velocity field than
the low ionization BELs. The ULQSOs FWHM tends to indicate that the
putative planar flow has a much larger random component than
previously acknowledged and the planarity of the low ionization BLR
has been overstated in the literature, since it was shown in section
3 that $c_{1}/c_{2}\approx 1$ in equation (2.2) (from the methods of
\citet{lab06} this implies that the isotropic gas and the planar
flow are equal contributors to the FWHM). Formally, these results
apply only to radio loud quasars. One might be suspicious that the
results presented are skewed by sample selection biases. First of
all, the anomalously large axial velocity component in the context
of the equatorial low ionization BLR model is corroborated by three
control samples, 2 of which are virtually complete. There does not
appear to be any obvious selection bias in choosing ULQSOs. It was
noted in section 3 that a selection bias does not appear to exist
based on the consistency with the carefully matched samples in
\citet{jar06}. The large redshifted BEL velocities in ULQSOs is
concluded to be robust and the dynamics of this phenomenon requires
further theoretical investigation. Any deductions beyond this are
speculative due to the large scatter in the data.

\clearpage
\begin{table}
\caption{\footnotesize{Broad Line Widths of Ultraluminal Quasars
\tablenotemark{a}}} {\footnotesize
\begin{tabular}{cccccccc} \tableline \rule{0mm}{3mm}
 Source &   z & max $\beta_{a}$& CIV FWHM &  CIII] FWHM &  MgII FWHM   & H$\beta$ FWHM  &  ref\tablenotemark{b}\\
  &    & & (km/s) &  (km/s)  & (km/s)  & (km/s) &  \\
\tableline \rule{0mm}{1mm}
 0016+731      & 1.781 &  11.9 & 5610  & 5390 &  4172   & ... &  1, 4 \\
 0106+013      & 2.107 &  23.3 &  5100 & 5700 &  ...   & ... &  1, 5 \\
 0332+321      & 1.263 &  24.5 &  4000  & 5700 &  2900   & ... & 1, 6 \\
 0336-019      & 0.852 &  12.4 & ...  & ... &  ...   & 4875 &  2, 7 \\
 0420-014      & 0.915 &  14.2 & ...  &  ... & ...   & 3480 &  1, 8 \\
 0736+017      & 0.191 &  11.5 & ...  & ... & ...   & 4628 &  1, 7 \\
 0827+243      & 0.939 &  27   &  6099 &  ... & ...   & ... &  2, 9 \\
 0836+710      & 2.16 &   14.1 &  6775  & 10403 &  3047   & 3410 & 2, 4, 10\\
 0850+581      & 1.322 &  12.7 &  ...  & 5107 & 6407   & ... &  1, 4 \\
 0859-140      & 1.327 &  16.3 &  4500 & 5400 &  4400   & 5700 &  1, 6, 11 \\
 0906+015      & 1.108 &  11.8 &  ... & 6160  &  3990   & ... &  1, 8 \\
 0945+408      & 1.252 &  22.5 &  3454 & 6003 &  7593   & ... &  1, 4 \\
 0953+254      & 0.712 &  12.4 &   ... & ... &  ...   & 3990 &  1, 7\\
 1055+201      & 1.11 &  10    & 3698 & 2979 &  ...   & ... &  1, 9 \\
 1127-145      & 1.187 &  19.8 & 4880 & 6190 &  2750   & ... &  2, 8 \\
 1156+295      & 0.729 &  14.1 & 5072  & 5601 &  2664  & 3700 &  1, 9, 12 \\
 1226+023      & 0.158 &  14.1 & 3941 & ... &  ...   & 4040 &  2, 13, 14 \\
 1253-055      & 0.536 &  10.4 & 5502  & 2769 &  4037   & 3100 & 3, 15, 16 \\
 1308+326      & 0.996 &  16.3 & ... & ... &  4016   & ... &  1, 17 \\
 1406-076      & 1.493 &  28.4 & ...  & 5800 &  3650   & ... &  2, 8 \\
 1508-055      & 1.191 &  31.4 & 1680  & 2160 &  2230   & ... &  1, 8 \\
 1510-089      & 0.36 &   18.9 & ... & ... &  2680  & 2796 &  1, 8, 13 \\
 1611+343      & 1.401 &  24 & 4744  & 5575 & ...   & 5600 &  1, 11, 16 \\
 1633+382      & 1.814 &  11.5 & 4332 & 4840  &  9650   & ... &  2, 4 \\
 1641+399      & 0.594 &  17 &   4829 & 4098 &  4196   & 4320 &  1, 4, 16 \\
 1642+690      & 0.751 &  16 &  ...  & 3190 &  4896  & 1850 &  1, 4 \\
 1655+077      & 0.621 &  15.3 &  ...  & ... &  2080   & ... &  1, 8 \\
 1828+487      & 0.691 &  14.8 & .... & ...  &  3010   & 4245 &  1, 4, 18 \\
 1901+319      & 0.635 &  15.2 & ...  & ... &  ...   & 2580 &  2, 19\\
 2223-052      & 1.404 &  32.5 & 3449  & 4418 &  ...   & ... &  1, 16 \\
 2243-123      & 0.63 &  10.7 & ...  & ... & ...   & 4258 &  1, 18 \\
 2230+114      & 1.032 &  11.7 & 3340  & 3903 &  ...   & ... &  2, 16 \\
 2251+158      & 0.859 &  15.7 & 3145  & 5175 &  ...   & 3700 &  2, 7, 16 \\

\tableline \rule{0mm}{1mm}
\end{tabular}}
\tablenotetext{a}{Column (1) is the source, followed by the redshift
and the maximum $\beta_{a}$ in the three studies. Columns (4), (5),
(6) and (7) are the FWHM of CIV, CIII], MgII and H$\beta$,
respectively. Column (8) is a list of the references, the first
reference is to the VLBI data and the remainder are for the FWHM.}
\tablenotetext{b}{References: 1. \citet{kel04} 2. \citet{jor01} 3.
\citet{ver94} 4. \citet{law96} 5. \citet{cor94} 6. \citet{bro94} 7.
\citet{jac91} 8. \citet{bro86} 9. \citet{kur04} 10. \citet{mci99}
11. \citet{net95} 12. \citet{bro96} 13. \citet{osh02} 14.
\citet{ves06} 15. \citet{wil86} 16. \citet{wil95} 17. \citet{wan04}
18. \citet{mar03} 19. \citet{gel94}}
\end{table}
\clearpage

\clearpage
\begin{table}
\caption{\footnotesize{Broad Line Widths of Ultraluminal Quasars
versus Steep Spectrum Radio Loud QSOs \tablenotemark{a}}}
{\footnotesize
\begin{tabular}{cccccccccc} \tableline \rule{0mm}{3mm}
 Line &    & ULQSOs &  &    &   & & SSQSRs & & \\
 \tableline \rule{0mm}{1mm}
  &  FWHM & median  & z &   number & FWHM  & median & z & number & ref\tablenotemark{b}  \\
  &  (km/s)  & &  &   & (km/s)  &  & &  &\\
\tableline \rule{0mm}{1mm}
H$\beta$      & 3898 &  3990 & 0.80  & 17 &  5460   & 4500 & 0.59 & 29 & 1 \\
      & $\pm 1011 $ &   & $\pm 0.50$  &  &  $\pm 3639 $   &  & $\pm 0.39$ & & \\
 H$\beta$      & 3898 &  3990 & 0.80  & 17 &  5836   & 5530 & 0.51 & 52 & 1, 2 \\
      & $\pm 1011 $ &   & $\pm 0.50$  &  &  $\pm 3476 $   &  & $\pm 0.32$ & & \\
 MgII     & 4125 &  3990 &  1.11 & 19 &  6398   & 6222 & 0.98  &  24 & 3CRR 3, 4, 5, 6 \\
     & $\pm 1997$ &   &  $\pm 0.52$ &  &  $\pm 2668 $   &  & $\pm 0.39$  &  &  \\
 CIII]     & 5074 &  5390 &  1.26  & 21 &  3904   & 3700 & 1.59 & 43 & 6, 7 \\
     & $\pm 1707 $ &   &  $\pm 0.45 $  &  &  $\pm 1698 $   &  & $\pm 0.68$ &  & \\
CIV    & 4429 &  4500 &  1.27 & 19 &  4279   & 3900 & 1.60 & 45 & 6, 7\\
     & $\pm 1189 $ &   &  $\pm 0.47 $  &  &  $\pm 1369 $   &  & $\pm 0.68$ &  & \\

\tableline \rule{0mm}{1mm}
\end{tabular}}
\tablenotetext{a}{The first column in table 2 is the BEL under
consideration. Next, we list the ULQSO mean and median values of the
FWHMs from table 1 followed by the number of objects in the
subsample in columns (2)-(5). Columns (6)-(9) are the same for the
samples of SSQSRs.} \tablenotetext{b}{References: 1. \citet{guu01}
2. \citet{cor97} 3. \citet{law96} 4. \citet{aar05} 5. \citet{bro94}
6. \citet{wil95} 7. \citet{cor94}}
\end{table}
\clearpage


\begin{thebibliography}{}
\bibitem[Aars et al(2005)]{aar05}Aars et al 2005, AJ
  \textbf{130} 23
\bibitem[Brotherton (1996)]{bro96}Brotherton, M. 1996, ApJS \textbf{102}
1
\bibitem[Brotherton et al(1994)]{bro94}Brotherton, M. et al 1994, ApJ \textbf{423}
131
\bibitem[Corbin and Francis (1994)]{cor94} Corbin, M. and Francis, P. 1994, AJ
\textbf{108} 2016
\bibitem[Corbin (1997)]{cor97} Corbin, M. 1997, ApJS \textbf{113}
245
\bibitem[Elvis(2000)]{elv00} Elvis, M. 2000 ApJ \textbf{545} 63
\bibitem[Gelderman and Whittle (1994)]{gel94}Gelderman, R. and Whittle, M. ApJS
\textbf{91} 491
\bibitem[Gu et al(2001)]{guu01} Gu, M., Cao, X., Jiang, D.R. 2001, MNRAS \textbf{327}
111
\bibitem[Jackson and Browne(1991)]{jac91} Jackson, N., Browne, I.W.A. 1991 MNRAS \textbf{250}
422
\bibitem[Jarvis and McClure(2006)]{jar06}Jarvis, M., McClure, R.
2006, to appear in MNRAS astro-ph/0603231
\bibitem[Jorstad et al (2001)]{jor01} Jorstad, S. 2001 ApJS \textbf{134}
181
\bibitem[Kellermann et al (2004)]{kel04} Kellerman, K.I. et al (2004) ApJ \textbf{609}
539
\bibitem[Krolik(2001)]{kro01}Krolik, J. 2001, ApJ \textbf{551} 72
\bibitem[Kuraszkiewicz et al (2004)]{kur04} Kuraszkiewicz, J. et al 2004 ApJS \textbf{150}
165
\bibitem[Labita et al(2006)]{lab06}Labita, M., Treves, A., Falomo, R., Uslenghi, M. 2006, to appear in
MNRAS astro-ph/0609185
\bibitem[Lamastra et al(2006)]{lam06}Lamastra, A., Matt, G., Perola, C. 2006, to appear in
A \& A astro-ph/0609355
\bibitem[Lawrence et al.(1996)]{law96} Lawrence, C. et al 1996, ApJS \textbf{107}
541
\bibitem[Lind and Blandford(1985)]{lin85}Lind, K., Blandford, R.
1985, ApJ \textbf{295} 358
\bibitem[Marziani et al (2003)]{mar03} Marziani, P. et al 2003 ApJS \textbf{145}
199
\bibitem[McClure and Dunlop(2001)]{mcc01}McClure, R., Dunlop, J. 2001, MNRAS
  \textbf{327} 199
\bibitem[McClure and Jarvis(2002)]{mcc02}McClure, R., Jarvis, M. 2002, MNRAS
  \textbf{337} 109
\bibitem[McIntosh et al (1999)]{mci99} McIntosh et al 1999 ApJ \textbf{517}
73
\bibitem[Netzer et al(1995)]{net95} Netzer, H. et al 1995, ApJ \textbf{448}
27
\bibitem[Nikolajuk et al(2006)]{nik06} Nikolajuk, M., Czerny, B., Ziolkowski, J. and Gierlinski, M. 2006, MNRAS \textbf{370}
1534
\bibitem[Ogle et al(2006)]{ogl06} Ogle, P., Whysong, D. and Antonucci, R. 2006
astro-ph/0601485
\bibitem[Oshlack et al(2002)]{osh02} Oshlack, A,, Webster, R., Whitting, M.
2002 ApJ \textbf{576} 81
\bibitem[Smith et al(2005)]{smi05}Smith, J., Robinson, A., Young, S., Axon, D., Corbett, E. 2005, MNRAS \textbf{359}
846
\bibitem[Vermeulen and Cohen(1994)]{ver94} Vermeulen, R . Cohen, M. 1994 APJ \textbf{430} 467
\bibitem[Vestergaard and Peterson(2006)]{ves06}Vestergaard, M. and Peterson, B. 2006, ApJ in press
astro-ph/0601303
\bibitem[Wang et al.(2004)]{wan04}Wang, J.-M., Luo, B, Ho, L. 2004, ApJL \textbf{615}
9
\bibitem[Wilkes (1986)]{wil86} Wilkes, B.J. 1986 MNRAS \textbf{218}
331
\bibitem[Wills and Brotherton(1995)]{wbr95} Wills, B., Brotherton, M 1995, ApJL
 \textbf{448} 81
\bibitem[Wills et al.(1995)]{wil95} Wills, B. et al 1995, ApJ \textbf{437}
139
\bibitem[Wills and Browne (1986)]{bro86} Wills, B.J., Browne, I.W.A. 1986 ApJ \textbf{302}
56
\end{thebibliography}
\end{document}